\documentclass[twoside]{article}
\usepackage{fleqn,espcrc2,amssymb}

\usepackage{graphicx}

\newcommand{\AmS}{{\protect\the\textfont2
  A\kern-.1667em\lower.5ex\hbox{M}\kern-.125emS}}
\title{Resistivity and electron-phonon coupling in YNi$_2$B$_2$C single crystals}

\author{R.S. Gonnelli\address{INFM - Dipartimento di Fisica, Politecnico di Torino, c.so Duca degli
Abruzzi 24, 10129 Torino, Italy},
         V.A. Stepanov\address{P.N. Lebedev Physical Institute, Russian Academy of Sciences,
         SU-117924 Moscow, Russia},
         A. Morello$^{\rm a}$,
         G.A. Ummarino$^{\rm a}$,
         G. Behr\address{Institut f\"{u}r Festk\"{o}rper- und Werkstofforschung Dresden, Postfach 270016, D-01171 Dresden, Germany},
         G. Graw$^{\rm c}$,
         S.V. Shulga$^{\rm c}$
         and
         S.-L. Drechsler$^{\rm c}$}

\begin{document}

\begin{abstract}
In this work, we present the results of precise measurements of
the resistivity of YNi$_2$B$_2$C single crystals with $T_{\rm c}$
= 15.5 K as a function of the temperature, and analyze the
experimental data in the framework of the Bloch-Gr\"{u}neisen
theory and electron-phonon coupling. The transport electron-phonon
spectral function that best fits the resistivity data is then
inserted in the real-axis Eliashberg equations, which are directly
solved to determine the normalized tunneling conductance both in
$s$- and $d$-wave symmetry. \vspace{-3mm}
\end{abstract}

\maketitle

\section{INTRODUCTION}\vspace{-1mm} The discovery of superconducting
quaternary rare-earth borocarbide intermetallic compounds
$R$Ni$_2$B$_2$C ($R$=rare earth) have aroused great interest
because, even though they are layered materials like high-$T_{\rm
c}$ cuprates, band-structure calculations predict a
three-dimensional electronic structure. Most of the experimental
and theoretical results tend to support a conventional
BCS-Eliashberg descriptions of the superconducting properties in
these materials \cite{ref1}. Some peculiar features of the upper
critical field of YNi$_2$B$_2$C have been recently explained in
the framework of Migdal-Eliashberg theory by considering the
presence of two bands, one of them being more deeply involved in
the transport properties of the compound \cite{ref2}.

In the present work we show that the resistivity measurements in
YNi$_2$B$_2$C are in complete agreement with the predictions of
the theory for the strong electron-phonon (e-p) interaction, and
we obtain a value of the transport e-p coupling constant in
agreement with previous experimental and theoretical results.
\vspace{-3mm}

\section{EXPERIMENT}\vspace{-1mm}
The resistivity of YNi$_2$B$_2$C single crystals with $T_{\rm c} =
15.5\,$K has been accurately measured as a function of the
temperature by using the standard four-probe technique. The
crystals were grown by using the rf - zone melting process
\cite{ref3}. We directly soldered the current leads to the lateral
sides of the samples, while very small gold voltage leads were
glued by a conducting paste to their surface. In order to improve
the sensitivity of the measure we injected in the crystals an AC
current (typically 10 mA at 133 Hz) and detected the voltage by
the standard lock-in technique. Figure 1 shows the resistivity
$\rho(T)$ of one of the YNi$_2$B$_2$C crystals. It is important to
notice that a very slow cooling down procedure allowed us to
collect nearly three thousand resistivity values between 4.2 and
300~K, but only few points are reported in the figure. As observed
in previous experiments \cite{ref4}, the resistivity shows a {\it
perfect} Bloch-Gr\"{u}neisen (BG)
behaviour with a linear high-temperature part ($\mathrm{d}\rho%
/\mathrm{d}T$ =~0.12 $\mu \Omega \cdot $cm/K) and a small residual
value $\rho(0)$=~3 $\mu\Omega \cdot $cm, indicating the high
quality and low impurity content of the samples. We obtained quite
similar results in various YNi$_2$B$_2$C samples.

\begin{figure}[t]
\vspace{-4mm}
\includegraphics[keepaspectratio,width=7.5cm]{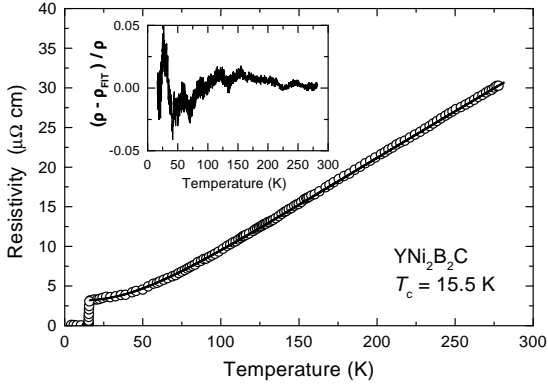}
\vspace{-13mm} \caption{\small{Few points of the typical
resistivity of one of our YNi$_2$B$_2$C single crystals (open
circles) and the fit using the Bloch-Gr\"{u}neisen model
(continuous line). The inset shows the relative deviations
\mbox{$(\rho\!-\!\rho _{\rm{FIT}})\!/\!\rho$} of the experimental
data from the fit.}} \vspace{-6mm}
\end{figure}

\vspace{-3mm}
\section{DISCUSSION}\vspace{-1mm}
According to the Matthiessen's rule for the resistivity in a
normal Fermi-liquid metal, one can write: $\rho (T)$=$\rho
_0$+$\rho _{\rm{ph}}(T)$, $\rho _0$ and $\rho _{\rm{ph}}(T)$ being
the residual and the phonon resistivities, respectively. In the
framework of the BG theory, the high-temperature part of $\rho%
_{\rm{ph}}(T)$ is well represented by a linear behaviour: $\rho%
_{\rm{ph}}(T)$=$(2\pi \varepsilon _{0}k_{\rm{B}}/\hbar \omega
_{\rm{p}}^{2})\lambda _{\rm{tr}}T$ where
$\lambda_{\rm{tr}}$=$2\int_0^\infty [\alpha_{\rm{tr}}^2 F(\Omega%
)/\Omega ]d\Omega$ is the transport \mbox{e-p} coupling constant, $\omega%
_{\rm{p}}$ is the plasma frequency and $\alpha_{\rm{tr}}^2%
F(\Omega )$ is the transport e-p spectral function. From the
linear part of the resistivity of Fig.~1 ($T\!\!>\!\!$ 100~K) and
the experimental value of the plasma energy
$\hbar\omega_p\!\!=\!\!4.25$ eV determined by means of reflectance
and EELS measurements \cite{ref5}, we extracted the transport
coupling constant $\lambda_{\rm{tr}}\!=\! 0.53$.

In order to obtain additional information on the \mbox{e-p}
coupling in YNi$_2$B$_2$C, we can fit the resistivity of Fig.~1 in
the whole temperature range by using the most general expression
for $\rho_{\mathrm{ph}}(T)$, according to the BG theory:
\vspace{-2mm}
\[
\rho_{\rm{ph}}(T)=(4\pi \varepsilon _{0}k_{\rm{B}} T/\hbar \omega
_{\rm{p}}^{2})\int_{0}^{\Omega _{\max }}[\alpha
_{\rm{tr}}^{2}F(\Omega )/\Omega ]
\]
\vspace{-6mm}
\[
\;\;\;\;\;\;\;\;\;\;\;\;\;\;\;\;\cdot[\hbar
\Omega/2k_{\rm{B}}T\sinh (\hbar \Omega /2k_{\rm{B}}T)]^{2}d\Omega.
\]
\vspace{-2mm}

Actually, we used for $\alpha_{\rm{tr}}^2 F(\Omega)$ the phonon
spectral density $G(\Omega)$ determined by inelastic neutron
scattering \cite{ref6} multiplied by a multistep weighting
function. As a first approximation, we considered a two-step
function, whose constant values for $\Omega\! <\! 37.5$ meV and
$37.5\!<\! \Omega\! <\!70$ meV (corresponding to the two
well-distinguishable structures of the $G(\Omega)$) were
determined by the fit. As shown in Fig.~1, the results of the fit
are {\it extremely good}: the theoretical BG curve fits so
perfectly the experimental $\rho (T)$ that the relative deviations
$(\rho -\rho _{\rm{FIT}})/\rho$ never exceed $\pm 5\%$ (see the
inset of Fig.~1). The spectral function $\alpha_{\rm{tr}}^2
F(\Omega )$ obtained from the fit is shown in the inset of Fig.~2.
Of course, the resulting $\lambda_{\rm{tr}}$ coincides with that
previously obtained from the linear part of $\rho (T)$.

As a first approximation, since $\lambda \approx
\lambda_{\rm{tr}}$, we used $\alpha_{\rm{tr}}^2 F(\Omega )$ to
calculate the YNi$_2$B$_2$C quasiparticle density of states by
directly solving the real-axis Eliashberg equations both in $s$-
and $d$-wave symmetry. The resulting tunneling conductances at
4.2~K are shown in Fig.~2. Actually a value $\lambda$=~0.57
slightly greater than $\lambda_{\rm{tr}}$ is necessary to explain
both $T_{\rm c}$ and the superconducting gap $\Delta\approx$ 2 meV
as determined in tunneling experiments \cite{ref7}. The comparison
of the curves of Fig.~2 with the experimental data \cite{ref7}
suggests a possible $s$-\emph{wave symmetry} for the
YNi$_2$B$_2$C. Finally, the value of $\lambda_{\rm{tr}}$  here
determined from the resistivity is consistent with the value used
in Ref.~2 for the discussion of the properties of YNi$_2$B$_2$C in
the framework of the isotropic single-band model.

\begin{figure}[t]
\vspace{-5mm}
\includegraphics[keepaspectratio,width=7.5cm]{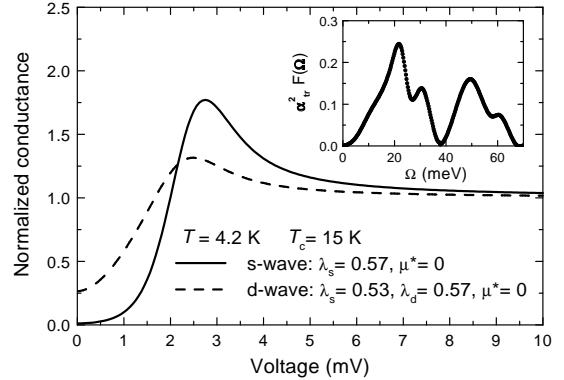}
\vspace{-12.5mm} \caption{\small{$s$- and $d$-wave tunneling
conductance of YNi$_2$B$_2$C calculated by direct solution of the
Eliashberg equations. In the inset the e-p spectral function
determined from the fit of the resistivity of Fig.~1 and used for
the conductance calculation is shown.}} \vspace{-6mm}
\end{figure}

\end{document}